\documentclass[manuscript]{aastex}

\usepackage{graphicx}

\shorttitle{Fluorine in the Sun and in OCs}
\shortauthors{Maiorca et al.}

\begin{document}


\title{A new solar fluorine abundance and a fluorine determination in the two open clusters M~67 and NGC~6404.}


\author{E. Maiorca\altaffilmark{1,2}, H. Uitenbroek\altaffilmark{3}, S. Uttenthaler\altaffilmark{4}, S. Randich\altaffilmark{1},  M. Busso\altaffilmark{2}, L. Magrini\altaffilmark{1}}


\altaffiltext{1}{INAF, Osservatorio Astrofisico di Arcetri, Largo E. Fermi 5, 50125 Firenze, Italy; maiorca@arcetri.astro.it}
\altaffiltext{2}{INFN, Sezione di Perugia; Via A. Pascoli, 06123 Perugia, Italy}
\altaffiltext{3}{AA(National Solar Observatory, Sunspot, USA), AB(National Solar Observatory, Sunspot, USA)}
\altaffiltext{4}{University of Vienna, Department of Astrophysics, T\"urkenschanzstra\ss e 17, 1180 Vienna, Austria}


\begin{abstract}
We present a new determination of the solar fluorine abundance together with 
abundance measurements of fluorine in two Galactic open clusters. We
analyzed a sunspot spectrum, observed by L. Wallace and
W. Livingston with the FTS at the McMath/Pierce Solar Telescope
situated on Kitt Peak and spectra of four giants in the old cluster
M~67 ($\sim$4.5 Gyr) and three giants in the young cluster NGC~6404
($\sim$0.5 Gyr), obtained with the CRIRES spectrograph at
VLT. Fluorine was measured through synthesis of the available HF
lines. We adopted the recent set of experimental molecular parameters
of HF delivered by the HITRAN database, and found a new solar fluorine
abundance of $A(F) = 4.40\pm 0.25$, in good agreement with the M~67
average fluorine abundance of $A(F) = 4.49\pm 0.20$. The new solar abundance
is in a very good agreement with the meteoritic value. 
The used modern spectrosynthesis tools, the agreement with the meteoritic value and with the results in open cluster M67, known to be a solar analogue, make our solar determination very robust. At the same time, the fluorine measurement in the
above-mentioned open clusters is the first step in the understanding of
its evolution during the last $\sim$10 Gyr in the Galactic disk. In order to develop this
project, a larger sample of open clusters is required, so that it
would allow us to trace the evolution of fluorine as a function of time
and, in turn, to better understand its origin. 

\end{abstract}





\section{Introduction}
The origin and the evolution of fluorine in the Galaxy are still nowadays a matter of debate. The available observational constraints, coupled with stellar nucleosynthesis models, have not yet clarified which stellar mass ranges and in which evolutionary stages are the mainly responsible for the fluorine production. Therefore, further and new observational evidence is needed, to understand where fluorine is produced and its implications on the stellar nucleosynthesis and Galactic chemical evolution.

The state of the art proposes three means of fluorine production: neutrino spallation on
$^{20}$Ne in gravitational supernovae (SNII; Woosley \& Haxton 1988),
hydrostatic nucleosynthesis in the He-burning core of heavily
mass-losing Wolf-Rayet (WR) stars (Meynet \& Arnould 2000), and
hydrostatic nucleosynthesis in the He-rich intershell of thermally
pulsing (TP) AGB stars (Busso et
al. 1999). It is still unknown which of the three above sources is the
main contributor for fluorine.

The above scenario is based on several observational studies that, during the last decade, have been
addressed the problem of the fluorine origin and evolution. Fluorine determinations were carried out
in different environments: namely: i) in the Large Magellanic Cloud (LMC) \citep{cun03};
ii) the globular cluster M~4 \citep{smi05} and M 22 \citep{dor13}; iii) the Milky Way
Bulge \citep{cun08}; iv) pre-main sequence stars of the Orion nebula cluster
\citep{cun05} and dwarf stars of the solar neighbourhood \citep{rec12}; v) Galactic and extragalactic asymptotic giant branch (AGB) stars \citep{abi09,abi10,abi11,utt08};
vi) in one hot post-AGB star \citep{wer05}; vii) in C-Rich
low-metallicity stars \citep{luc11}; viii) in planetary nebulae
\citep{zha05}; ix) in the interstellar medium surrounding Type II supernovae \citep{fed05}. These recent studies enlarged and in some cases
reanalyzed the sample of stars presented in \citet{jor92}.

Almost all the above fluorine analyses have been developed using
spectral features of the HF molecule (mostly the R 9 line at
$\lambda_{vacuum}$=2336.47 nm). A \textit{theoretical} list of HF
molecular parameters (e.g. log\textit{gf}, E$_{low}$), provided by
R.H.\ Tipping (see e.g.\ Abia et al.\ 2009), was in general adopted, together
with an old solar abundance derived by \citet{hal69}, A(F)$_{\odot}$=4.56. Very recently, a new list of \textit{experimental}
molecular parameters for the HF
molecule has been delivered by the HITRAN database
\citep[see][for details on this database]{rot13}.
Therefore, we started a new analysis of the fluorine abundance based on these new
data. More in detail: i)
we reanalyzed the solar fluorine abundance as observed in sunspot
spectra with modern techniques: sunspot specific spectrosynthesis
simulations and atmospheric model. ii) We collected, for the first
time, spectra of giant star members of two Galactic open clusters
(OCs), M~67 and NGC~6404, in the infrared region, where HF
lines were detected and analyzed.

The solar fluorine abundance is used as a zero-point for all the other
dedicated studies, hence a redetermination in light of the above
new molecular parameters and of more recent analysis techniques was
needed \citep[see also][]{asp09}.

On the other hand, this work is the starting point of a new project which
consists in the fluorine determination in several open clusters, with
different ages and Galactocentric distances (R$_{GC}$). In fact, this
investigation offers the opportunity of measuring fluorine evolution
during the last $\sim$10 Gyr as a function of time. This can be done
since the age estimate of OCs can be performed with a smaller
uncertainty than for field stars. In turn, the
knowledge of the fluorine evolution provides also a further constraint to
understand which kind of stars (low mass or more massive stars) is
mostly responsible to its production. At the same time, the
analysis of the M~67 cluster is also a good test of the
solar fluorine determination, since this cluster shows a solar-like
abundance distribution and its age, metallicity and R$_{GC}$ resemble
those of the Sun, so that it can be considered as a \textit{solar
  analogue} cluster.

In Section 2 we describe the observations and the analysis for the fluorine determination in the Sun, while Section 3 is focused on fluorine in open clusters. Section
4 shows our results, while in Section 5 and 6 we discuss results,
giving our final conclusions.

\section{Observations and Analysis in the Sun}
\subsection{Umbral atlas\label{sec:umbralatlas}}
In order to determine the solar fluorine abundance we employed the spectral atlas
by \citet{wal01} of a medium strong sunspot umbra, observed on
1982 May 16 with the Kitt Peak FTS near disk center ($\mu = 0.996$).
The spectral atlas has a resolution of $\lambda/\Delta \lambda =
480,000$. The observed spot had an associated magnetic field strength
of 2490 G. We used spectral lines of the OH (in the spectral region
around 1.565 $\mu$m) and CO molecules (near the HF lines we analyzed)
to decide on the appropriate effective temperature of models to
be used for the fluorine abundance determination. The umbral atlas is
corrected for telluric absorption, but not for scattered light
originating from the much brighter surrounding photosphere.
However, in the infrared the contribution from scattered photospheric
light is only of the order of a few percent in the continuum
\citep[see][]{van84}, and should have little effect on the
shallow HF lines we employ.

\subsection{Tools, molecular parameters, linelist}
We determined the abundances of F by fitting the observed
spectra with simulated ones.

For fitting the solar umbral atlas we used the RH code of 
\citet{uit00,uit01,uit04} to calculate molecular
spectral lines in LTE from the
one-dimensional radiative equilibrium models of \citet{kur93}
with different effective temperatures. The transfer code was
used to solve chemical equilibrium for the most abundant
molecules in the solar atmosphere, H$_{2}$,
C$_{2}$, N$_{2}$, O$_{2}$, CH, CO, NH, NO, OH, and the HF molecule.
Dissociation energies and parametrization with temperature
of the equilibrium constants and partition functions for these
molecules were taken from \citet{sau84}. Line lists for
the CO and OH molecules were taken from \citet{goo94} and Kurucz (\texttt{http://kurucz.harvard.edu/LINELISTS/LINESMOL/}), respectively.

Molecular parameters for HF lines have been recently measured and can
be found in the HITRAN database (\texttt{http://www.cfa.harvard.edu/hitran/}).
As we said in the Introduction, this
is a relevant improvement in fluorine studies, since up to now these
data were provided by R.H.\ Tipping through theoretical calculations.
The adopted HF
line list is shown in Table \ref{tab_flulin}.


\begin{table}[ht]
\begin{center}
\caption{Adopted HF line list \label{tab_flulin}}
\begin{tabular}{lccc}
\tableline\tableline
Line  &    Wavelength  & E$_{LOW}$          &  log gf \\
      &      (nm, in vacuum)      &  (eV)       &         \\
\tableline
R21      & 2270.898  &   1.124 (1.378) & -4.078 (-4.087)\\
R20      & 2270.175  &   1.026 (1.280) & -4.045 (-4.053)\\
R19      & 2270.569  &   0.932 (1.186) & -4.017 (-4.025)\\
R16      & 2278.447  &   0.674 (0.929) & -3.957 (-3.964)\\
R15      & 2283.310  &   0.597 (0.851) & -3.945 (-3.951)\\
R14      & 2289.298  &   0.524 (0.778) & -3.937 (-3.943)\\
R13      & 2295.792  &   0.455 (0.710) & -3.932 (-3.938)\\
R11      & 2313.473  &   0.332 (0.586) & -3.935 (-3.941)\\
R9       & 2336.470  &   0.227 (0.482) & -3.956 (-3.961)\\
R8       & 2348.803  &   0.182 (0.436) & -3.975 (-3.980)\\
R7       & 2362.997  &   0.142 (0.396) & -4.000 (-4.005)\\
R6       & 2378.434  &   0.107 (0.361) & -4.033 (-4.038)\\
R1       & 2475.206  &   0.005 (0.259) & -4.466 (-4.470)\\
\tableline
\end{tabular}
\tablenotetext{}{Values in parentheses are from the R.H.\ Tipping list.}

\end{center}
\end{table}

In the solar case we accounted
for Zeeman splitting in the molecular OH and HF lines 
\citep[see][]{ber02,uit04}
due to the 2490 G vertical magnetic field in the observed sunspot umbra.
In particular, since the ground state of the HF molecule is X $^1\Sigma$,
with a total orbital angular momentum $\Lambda = 0$ we could use
Hund's case (b) for weak or absent spin-orbital coupling to calculate
the line splittings by the Zeeman effect. For most HF lines
at the wavelengths of interest the effective Land\'e $g$ factors turned
out to be very small, around 0.03 to 0.05, except for the HF R1
line, which has a larger factor of 0.25. Thus, in general the Zeeman
effect in HF lines matters very little for the solar fluorine abundance
determination from an umbral spectrum. 

\subsection{Solar fluorine abundance determination}
\subsubsection{Determination of the Sunspot umbral effective temperature}
\begin{figure}[ht]
\includegraphics[width=\textwidth]{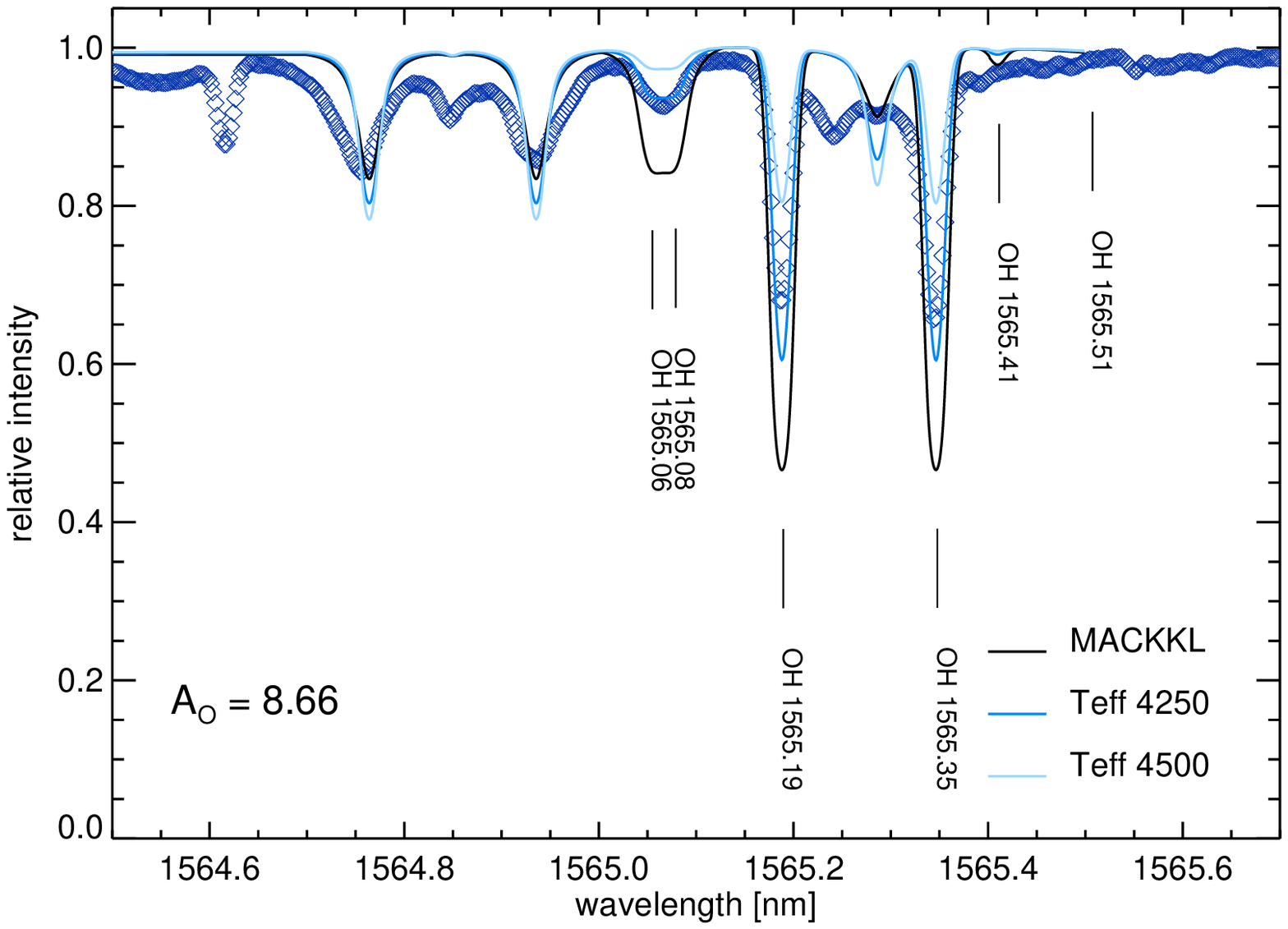}
\caption{Umbral spectrum (dark blue diamonds) and model spectra for
three 0ne-dimensional models: an umbral model (MACKKL, solid black),
and two radiative equilibrium models at effective temperatures
of 4250 K (solid medium blue) and 4500 K (solid light blue.
Line calculation includes the Zeeman effect for a 2500 G constant
vertical magnetic field.\label{fig_OH}}
\end{figure}

Since the association--dissociation equilibrium of the HF molecule
is strongly temperature-dependent we need to accurately determine
the effective temperature of the atmospheric model that is most compatible
with the solar umbral atlas we used (Section ~\ref{sec:umbralatlas}).
We accomplished this by matching CO lines in the range of the pertinent HF lines,
and OH lines in the 1.5$\mu$m wavelength range. In particular, the latter
lines are highly temperature sensitive as is clear in Figure\ \ref{fig_OH},
which shows the solar umbral atlas (blue diamonds) around 1565 nm,
together with three model spectra: the medium umbral model by
\citet[][hereafter MACKKL]{mal86},
and two radiative-equilibrium models (with solar
gravity) from \citet{kur93} at effective temperatures of 4500 K (solid,
light blue) and 4250 K (medium blue). In all cases a constant vertical
magnetic field of 2500 G was imposed, and the Zeeman splitting of the
OH lines as well as the \ion{Fe}{1} line at 1564.85 nm was accounted for.
Zeeman splitting under these conditions does not affect the stronger
OH lines at 1565.20 nm and 1565.35 nm much because they have small effective
Land\'e $g$ factors of 0.08, while the blended OH lines at 1565.06 nm
and 1565.08 nm with Land\'e factors of 0.18 are slightly broadened.
The \ion{Fe}{1}
is completely split into its three components because of its large
Land\'e $g$ factor of 3. Note that the central $\pi$ component in the
model is weaker than in the observation because we assumed a vertical
field viewed at $\mu = 0.996$, while the actual (average) field was most likely
not as vertical, and that the aperture of the spectrometer was several
arcsec accross allowing it to sample different field strengths and
inclinations, explaining the much larger broadening of the iron line
spectral components compared to that of the OH lines. Clearly, the
$T_{\mathrm{eff}} = 4250$ K case provided the
best match to the OH lines, in particular for the weaker OH lines, which
form in the deeper layers of the umbral atmosphere, like the HF lines
we analyzed for our abundance determination. We, therefore, adopted this
latter model in what follows.

\subsubsection{Fluorine in the Sun}
\begin{figure}[p]
\includegraphics[width=\textwidth]{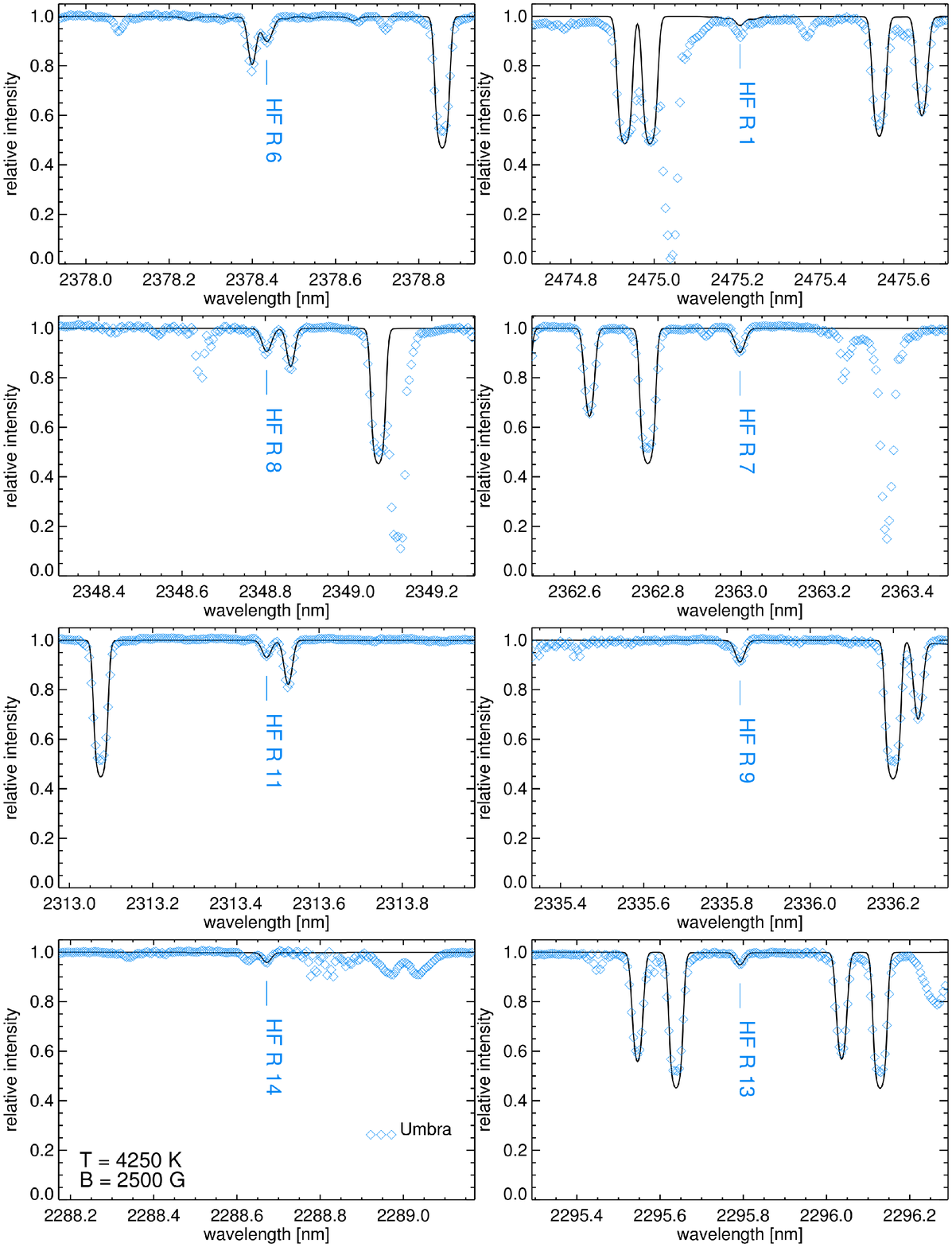}
\caption{Best match of HF lines in the 4250 K effective temperature
atmosphere (solid black) to the umbral atlas spectrum (blue diamonds).
Zeeman effect for a 2500 G constant vertical magnetic field is
accounted for in the HF lines.\label{fig_HF_sun}}
\end{figure}
We estimated the Solar fluorine abundance by determining a best
fit of the 8 HF feautures visible in the umbral atlas, namely
the R1 through R14 lines in the 2.2 -- 2.5 $\mu$m region of the spectrum
employing the $T_{\mathrm{eff}} = 4250$ K radiative equilibrium model
with a vertical magnetic field of 2500 G, constant with height.
Figure \ref{fig_HF_sun} shows the best fit with an abundance of
$A(F) = 4.40$. All HF lines are well reproduced with this value,
apart from the R1 line, which forms in a spectral range that is
heavily affected by telluric contamination, as are the regions
near 2475.05 nm, 2349.13 nm, 2363.35 nm, and 2288.8 nm. Other lines in the
8 spectral windows arise from the CO first overtone vibration--rotation
band. The weaker lines of this band, forming in similar layers
of the atmosphere as the HF lines are matched very well with
the $T_{\mathrm{eff}} = 4250$ K model atmosphere, indicating that our choice
for the effective temperature is appropriate. The stronger CO lines,
forming in higher layers are not so well matched, presumably because
these layers of the sunspot umbra may not be well described by
a hydrostatic radiative equilibrium model that does not account
for the structure of the spot's magnetic field.

The difference between the previously accepted value of the solar
fluorine abundance \citep[][, A(F)$_{\odot}$=4.56]{hal69} and our result stems mainly from our use of the
new experimental lower level energy values provided by
the HITRAN data base (see the differences in Table\ \ref{tab_flulin}).
Sources of uncertainty are discussed in Section \ref{sec_solaruncert}.
\subsubsection{Uncertainties in the Sun\label{sec_solaruncert}}
One of the largest uncertainties in the determined solar fluorine abundance
is the choice of abundances of oxygen and carbon we used, which influences
the choice of the effective temperature of the umbral model. We employed
the newer values of $A(O) = 8.66$ and $A(C) = 8.39$ recommended by
\citet{asp04} and \citet{asp05}, respectively. These values constitute
a downward revision from the values obtained with more traditional
one-dimensional models that is motivated by modeling in three-dimensional
simulations. Using the larger older values of $A(O) = 8.93$ and $A(C) = 8.60$
would force us to adopt a slightly hotter model for the umbra, mainly
because the OH lines at 1565 nm would strengthen too much in our canonical
model. With the $T_{\mathrm{eff}} = 4500$ K model the stronger OH lines
fit with the larger oxygen and carbon abundances, but the weaker ones are
too shallow. The first overtone CO lines near the HF lines are all slightly
too deep with the higher effective temperature and the larger abundances.
The fluorine abundance needed to fit the higher temperature model is
$A(F) = 4.65$, so that the uncertainty in effective temperature constitutes
an uncertainty in the fluorine abundance of about 0.25 dex. Overall, however,
the fits for OH, CO and HF lines is better with the smaller oxygen,
carbon and fluorine abundances and the 4250 K model. Moreover,
given the strongly reduced convection in sunspot umbrae, we feel
confident that the newer C and O abundances apply even when we use
static one-dimensional atmospheric modeling.

The other factor in the downward revision of the solar fluorine 
abundance we recommend is the new set of experimental HF line parameters from
the HITRAN data base. Indeed, as we said above, the uncertainty in the temperature estimate of the sunspot spectrum could affect the F abundance of about 0.25 dex. 
 If we use the old molecular parameters from Tipping, with the corresponding partition function in the spectrosynthesis code \citep[see][]{jon14}, and the C, O abundances from \citet{asp04,asp05},
we find a best match in the $T_{\mathrm{eff}} = 4250$ K model of 
$A(F) = 4.35$, slightly below the original determination of \citet{hal69}.

\section{Observations and Analysis in open clusters}
Our sample includes two Galactic open clusters, namely M~67 and
NGC~6404. The evolved members of which can be observed with a ground-based
8m-class telescope and are cool enough to allow the detection of HF in
their spectra.

M~67 is one of the most widely studied, best-known open clusters. Recent
works can be found in \citet{pac12,can11,ses05} (lithium abundance),
\citet{mai11,pan10,fri10,ran06} (elemental abundances), \citet{pas12}
(search for planets), \citet{cas11,bru14} (solar twins). Its age and elemental abundances are very close to the
solar values.

NGC 6404 has been studied by \citet{car05} with CCD photometry and
more recently by \citet{mag10}, who derived spectroscopic abundances
of Fe and $\alpha$-elements in four giants.  In Table
\ref{tab_clupar} we report the cluster parameters and references.

\begin{table}[ht]
\begin{center}
\caption{Open cluster parameters \label{tab_clupar}}
\begin{tabular}{rcccccc}
\tableline\tableline
OC  & E(B-V) & Age       & D$_{\odot}$  &R$_{GC}$  & [Fe/H] &   Ref. \\
    &         &    (Gyr)  & (kpc) &  (kpc)   &      &       \\
\tableline
M~67          &   0.05   & 4.3 & 0.908 & 8.639 & +0.03 &    a\\
NGC~6404      &    0.92  & 0.5 & 1.820     &   6.188     & +0.11 &  b     \\
\tableline
\end{tabular}
\tablenotetext{a}{\citet{ran06}}
\tablenotetext{b}{\citet{mag10}}
\tablenotetext{c}{R$_{GC\odot}$= 8 kpc}
\end{center}
\end{table}

\subsection{Target stars and data reduction}
We collected spectra of the OC stars with CRIRES, the Cryogenic Infra-Red
Echelle Spectrograph, mounted to the Nasmyth focus A at the 8.2 m Unit
Telescope No. 1 (Antu) of ESO’s VLT on Cerro Paranal, Chile.
Three giants in NGC~6404 and four giants in M~67 analyzed in this study were observed with CRIRES 
in 2012 (period 88). The analysis presented here refers to a wavelength
setting covering the range 2240-2295 nm (order 25). The slit width was
set to 0.2". The integration time has been calculated through the
CRIRES ETC for each target star. A hot standard star at similar air
mass was observed immediately afterward.

The raw frames were reduced with the CRIRES pipeline (ver. 2.2.1), and
the one-dimensional science and standard star spectra were
wavelength-calibrated separately using the numerous telluric
absorption lines present on all the four detector arrays. The
wavelength-calibration was done separately for the science and
telluric standard star spectra because of the limited reproducibility
of the Echelle grating position.  Finally, the science spectrum was
divided by the standard star spectrum to correct for the telluric
lines and the illumination pattern as well as possible. Note that the
telluric lines are strong enough to be used as wavelength calibrator,
but they are weak enough to be corrected for by standard star
division; thus they have no influence on the abundance measurements
presented here.

\begin{table}[ht]
\begin{center}
\caption{Log of observations  \label{tab_log}}
\begin{tabular}{lccc}
\tableline\tableline
Star\tablenotemark{a}  & Night of observation          &  Exp. Time (s) &  S/N (pixel)  \\
\tableline
\multicolumn{4}{c}{M~67}       \\          
S364      & 07 Mar 2012  &   60 &  59\\
S488      & 13 Jan 2012  &   30 & 108\\
S978      & 13 Jan 2012  &   60 &  89\\
S1250     & 13 Jan 2012  &   60 &  90\\
\tableline
\multicolumn{4}{c}{NGC~6404}       \\          
16        & 10 Mar 2012  &   240 & 109  \\
27        & 11 Mar 2012  &   240 & 123  \\
40        & 11 Mar 2012  &   360 & 109  \\
\tableline
\end{tabular}
\tablenotetext{a}{References for the star identification are \citet{san77} and \citet{car05} for M67 and NGC6404 respectively. For a useful cross-references table see the WEBDA database \citep{mer95}.}
\end{center}
\end{table}

\begin{figure}[ht]
\includegraphics[width=0.5\textwidth]{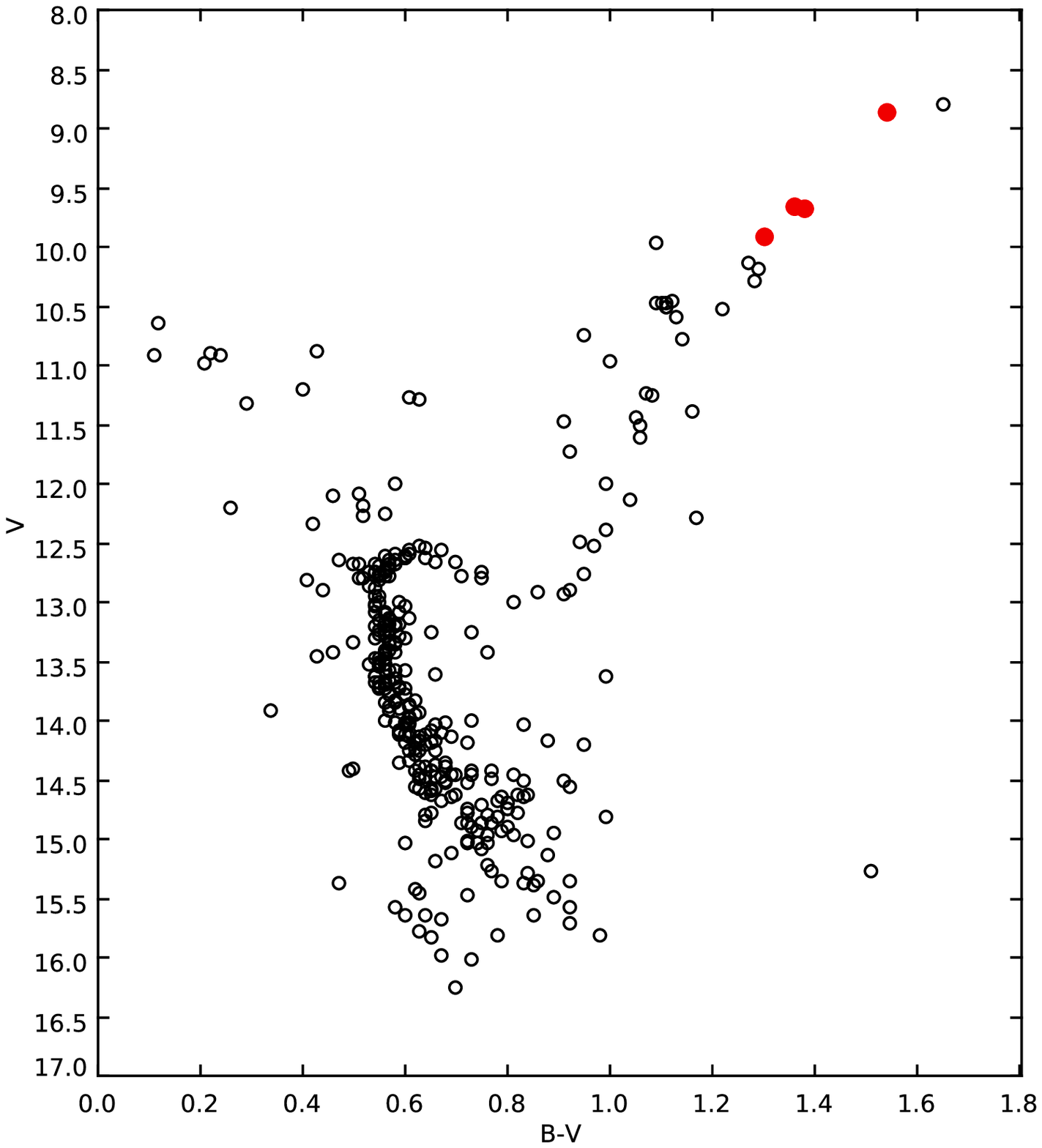}
\includegraphics[width=0.5\textwidth]{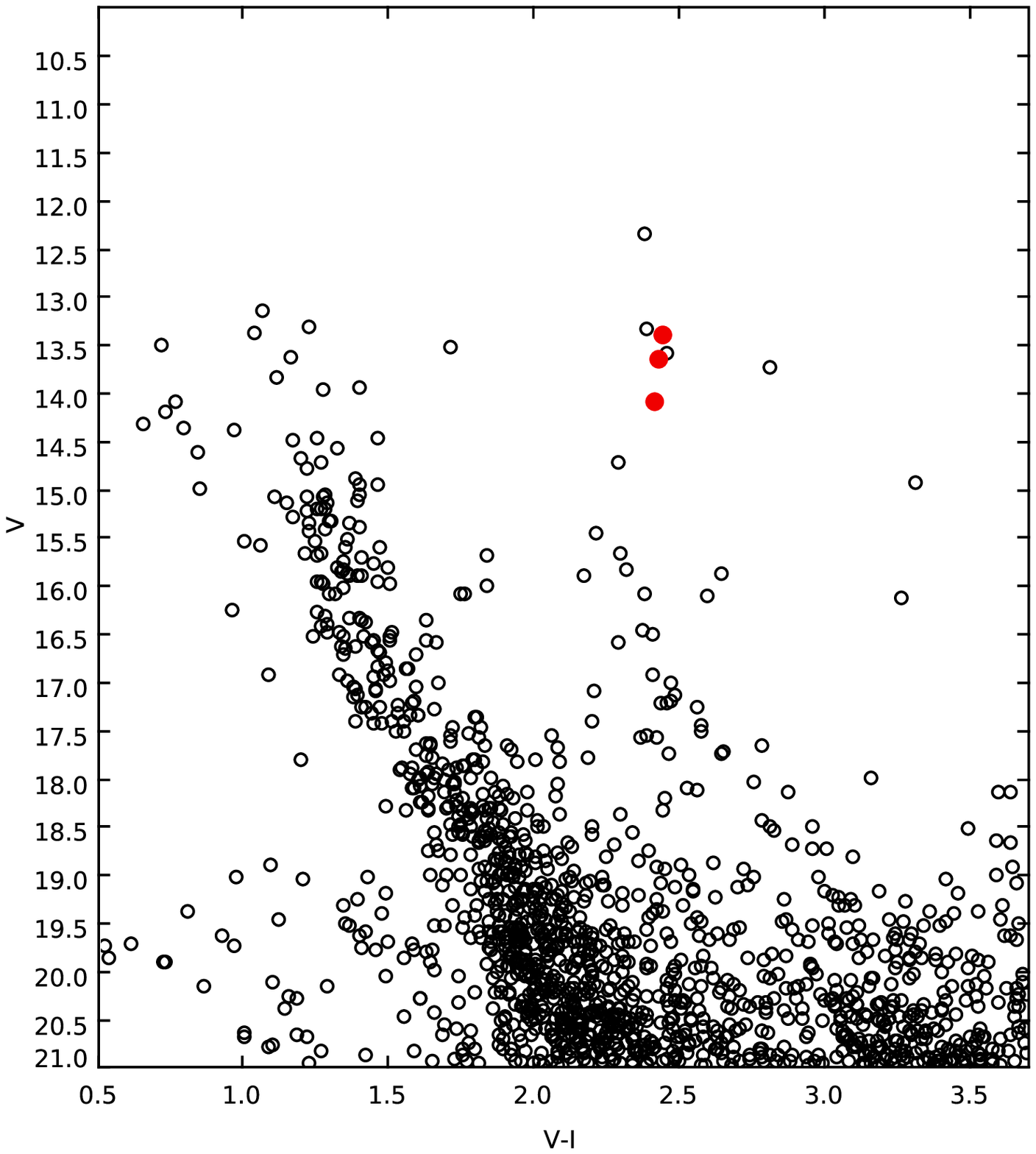}
\caption{CMD of M~67 (left panel) and of NGC~6404 (right
  panel). Red-filled circles represent our target stars. Photometric data were taken from the WEBDA database \citep{mer95}. \label{fig_CMD}}
\end{figure}

Table \ref{tab_log} shows the details of the observations, while in Table \ref{tab_pho} we show the available JHK photometry for the
observed stars, taken from the 2MASS catalogue \citep{skr06}. JHK values together with the reddening values for each cluster and the \citet{car89} extinction calibrations at
different bandpasses have been adopted to derive the (J-K)$_0$ values
of Table \ref{tab_pho}. In Figure \ref{fig_CMD} we show the CMD of the analyzed clusters, with
the target stars plotted in red. M~67 giants belong to different
stages of the RGB phase, while, according to \citet{mag10}, the three
giants of NGC~6404 are close to the RGB tip. According to current stellar evolutionary models \citep[see
  e.g.][]{cri09}, none of the stars in our sample should have produced or modified
fluorine by itself. Hence the F amount has been inherited from previous
generations of stars.

\textbf{Quite obviously, it is crucial to ascertain that the studied stars 
are real members of the cluster. \citet{mag10} verified
this through radial velocities. The membership is also supported by the metal-rich content of them found by previous authors ([Fe/H]=0.07,0.20,0.11 respectively). Moreover, \citet{car05}, in their Figure 8, overplotted an isochrone on the NGC6404 CMD. There one can see that our selected giants all lay on that isochrone. For all the above reasons we are confident about the membership of our analyzed giants.}
\begin{table}[ht]
\begin{center}
\caption{Sample stars photometry and stellar parameters\tablenotemark{a}  \label{tab_pho}}
\begin{tabular}{lccccccc}
\tableline\tableline
Star  & J          &  K &  (J-K)$_0$  &    T$_{eff}$  & log g        &  $\xi$        &  [Fe/H]       \\
      &            &    &               &      (K)      &   dex          &  (km s$^{-1}$)  &  dex     \\
\tableline
\multicolumn{8}{c}{M~67}      \\          
S364      & 7.542  &   6.690 & 0.825 & 4207 (4284)  &   1.91 (2.20) & 2.5 &    0.03 (-0.02)       \\
S488      & 6.010  &   5.010 & 0.963 & 3907  &   1.37 & 2.5 &    0.03  \\
S978      & 7.325  &   6.494 & 0.804 & 4255  &   2.01 & 2.5 &    0.03  \\
S1250     & 7.314  &   6.489 & 0.798 & 4269  &   2.03 & 2.5 &    0.03   \\
\tableline
\multicolumn{8}{c}{NGC~6404}      \\          
16        & 8.716  &   7.427 & 0.797  & 4273 (4450)  &   2.04 (1.65) & 2.5 (2.1)  &    0.11 (0.07)\\
27        & 8.989  &   7.685 & 0.812  & 4238 (4400)  &   1.97 (1.40) & 2.5 (1.8)  &    0.11 (0.20)\\
40        & 9.471  &   8.098 & 0.881  & 4084 (4250)  &   1.68 (2.30) & 2.5 (1.4)  &    0.11 (0.11)\\
\tableline
\end{tabular}
\tablenotetext{a}{Values in brackets are from \citet{bru14} for M 67 and from \citet{mag10} for NGC 6404 and were derived through spectroscopy.}
\end{center}
\end{table}

\subsection{Fluorine in open clusters}
As for the Sun, we determined abundances of F in OCs by fitting the observed
spectra with simulated ones.

We used the MOOG code by \citet{sne73}{, version 2010,} to perform the
spectrosynthesis. The synthetic spectrum is based on a one-dimensional 
LTE calculation. The stellar model atmospheres are
those provided by \citet{kur93}. Molecular equilibrium is solved for in the models
 and the Uns$\ddot{o}$ld approximation for the collisional broadening
is adopted. The linelist and the atomic/molecular parameters have been
provided by C.\ Sneden (private communication) and it is completed with
molecular CN and CO species. As for the Sun the HITRAN data were used for
the molecular line parameters of HF.

\subsubsection{Stellar parameters \label{stepar}}
We derived temperature, gravity and microturbulence for our stars,
from the relations available in \citet{leb12}. The mean metallicity of
the cluster was assumed as the metallicity for each star of the
corresponding cluster (references for the cluster metallicities are
reported in Table \ref{tab_clupar}). Our results are summarized in
Table \ref{tab_pho}.

\citet{mag10} derived stellar parameters for NGC~6404 stars from
spectroscopy, while no similar data are available for our M~67 target
stars, with the exception of the giant S364, for which recent spectroscopical derivation of its parameters can be found in \citet{bru14} (see Table \ref{tab_pho}). In order to be homogeneous, we adopted for all the sample stars
the parameters coming from photometry. In Section \ref{uncclu} we
will discuss the effect of a change in these values on the fluorine
abundance.

\subsubsection{CNO estimate \label{bleCNO}}
\begin{figure}[ht]
\includegraphics[width=\textwidth]{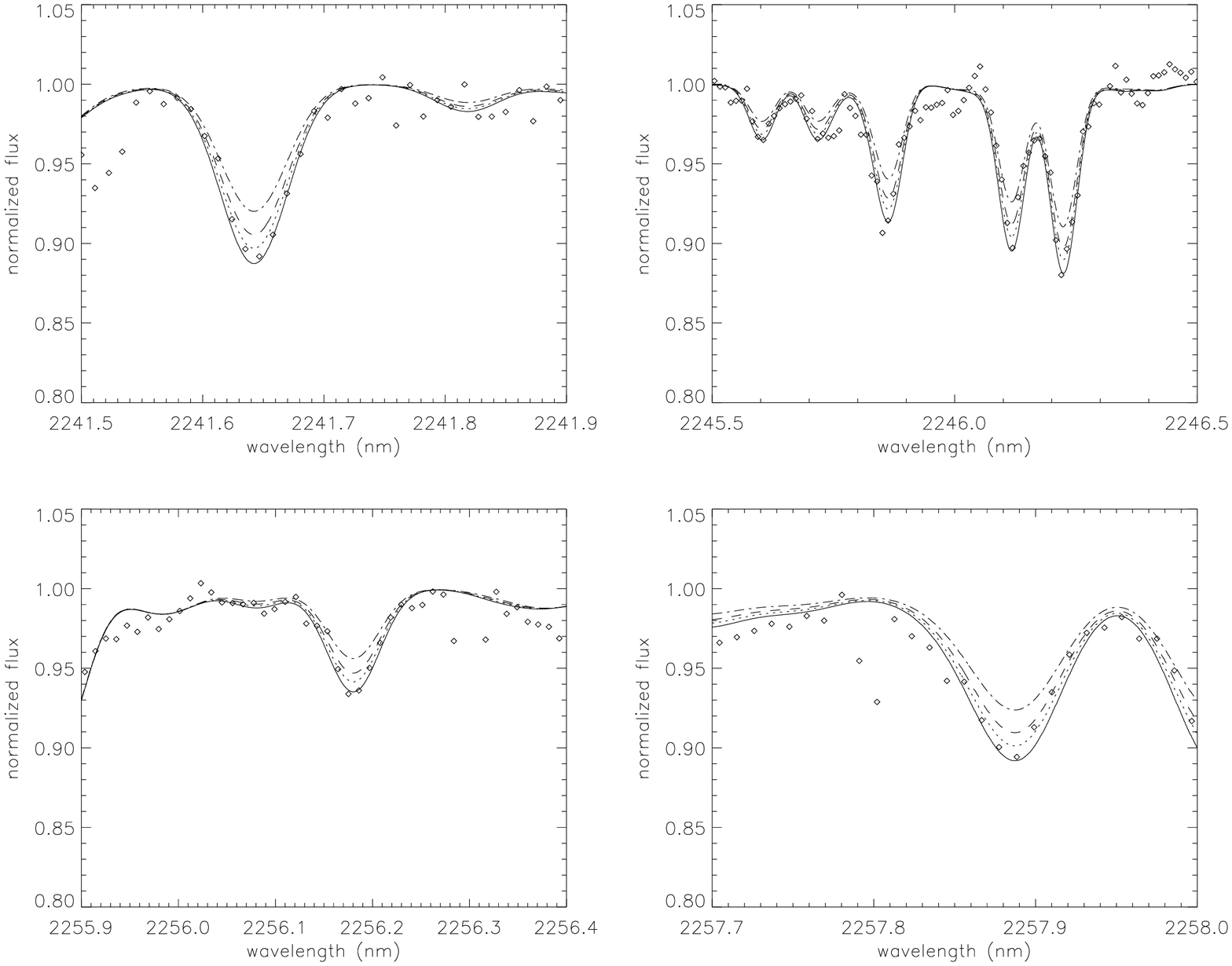}
\caption{Example of syntheses of CN for the M~67 giant S488. The
  solid line represents a synthetic spectrum calculated with a CN
  overabundance of +0.2 dex with respect to the dot-dashed
  simulation. \label{fig_bleCNO}}
\end{figure}

CN lines are present in the observed spectral range and all the
analyzed fluorine features in OCs are blended with them. We could not
retrieve an independent determination of the C, N and O abundances because 
CO lines are not detectable in our spectra. 
Therefore for
each star we used $\sim$30 CN lines, by which we found the total C+N+O
set of abundances that best fits the spectrum. The uncertainty on
the above sum is less than $\pm$0.1 dex. These estimates have been
adopted in the subsequent fluorine determinations, but we underline
that they do not provide information on the individual abundance
of C, N and O.  In Figure \ref{fig_bleCNO}, we show some examples of
synthesis for CN lines.

\subsubsection{Fluorine determination}
\begin{figure}[ht]
\includegraphics[width=0.5\textwidth]{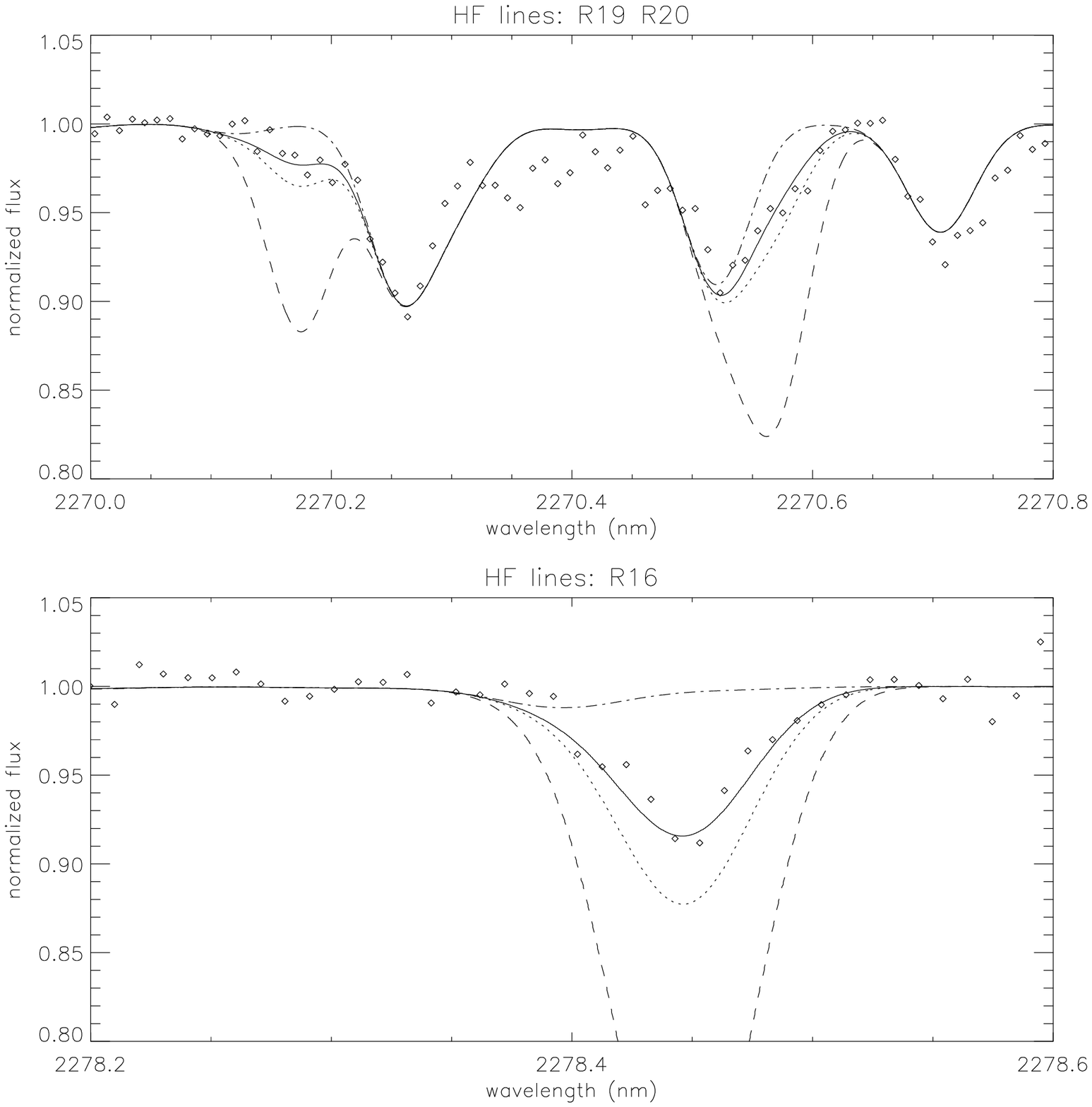}
\includegraphics[width=0.5\textwidth]{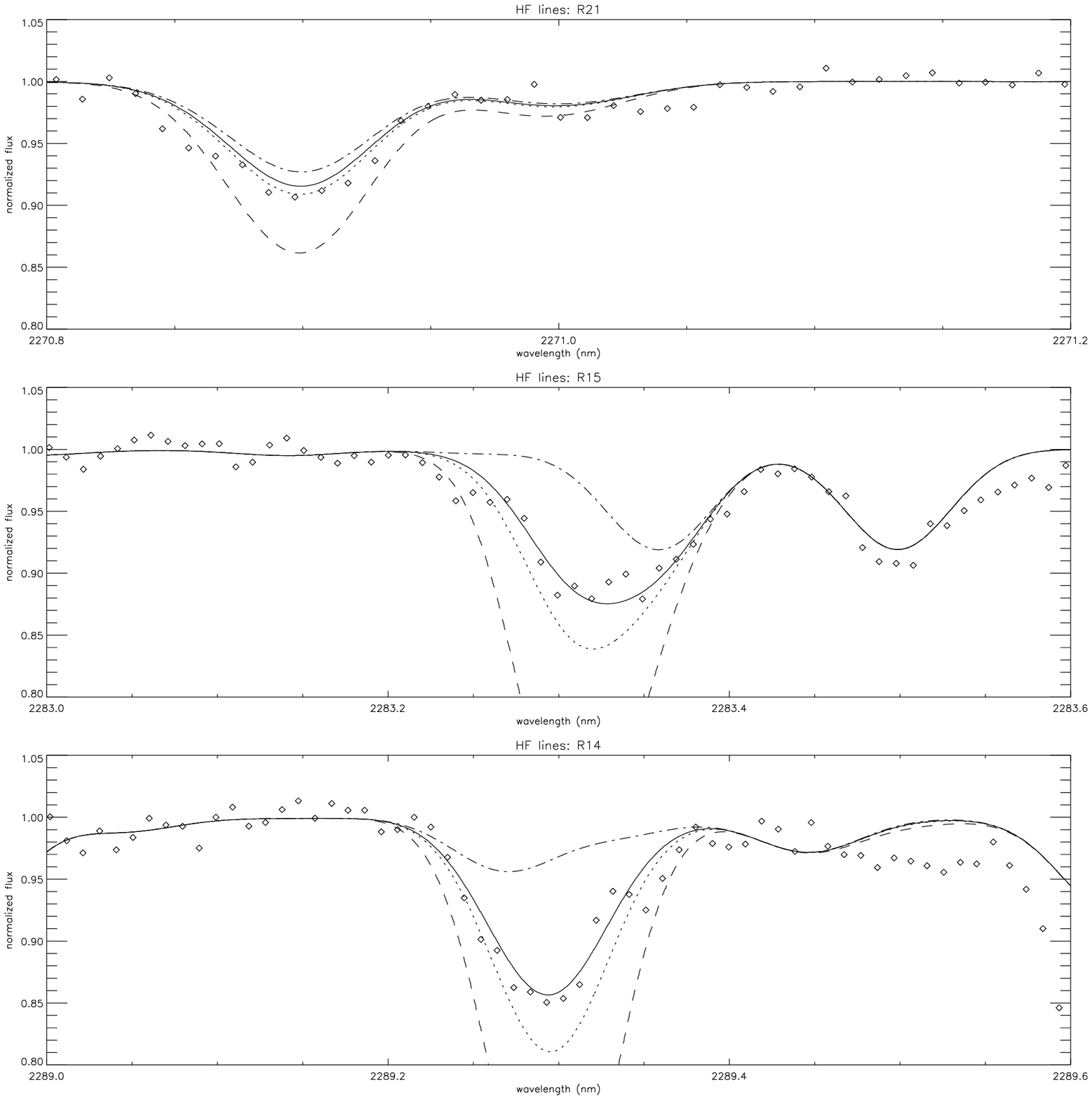}
\caption{\textbf{Comparison of observed and synthetic spectra around HF lines in the M~67 sample giant S488. We plotted the synthetic spectrum for the following choices. i) No fluorine (dot-dash line), in order to show the contribution of each CN blend to each HF line. ii) A(F)= 4.36 (solid line). iii) A(F)=4.56 (dot line) in order to show half of the range of our final uncertainty in the F abundance, this corresponds to the old solar fluorine abundance \citep{hal69}. iv) A(F)=5.16 } Left panels from top to bottom show HF lines R19, R20 and R16. Right panels from top to bottom show R21, R15 and R14  \label{fig_flufit}}
\end{figure}

Including the measured C+N+O values for each corresponding star, we
performed the HF synthesis. Figure \ref{fig_flufit} shows an example
of the synthesis for the six considered HF lines with four different
fluorine abundances in the M~67 giant S488.  The lower value
corresponds to A(F)=-1.56, while the upper value is equal to the old
solar abundance \citep{hal69}.  Analyzing the spectrum line-by-line in
Figure \ref{fig_flufit}, we can point out some features that remain
valid also for the rest of the studied stars: \textit{line R20} is
very weak and only marginally blended with the CN line at 2270.25
nm. \textit{Line R19} is partially blended with the CN line at 2270.52 nm,
but its contribution to the right wing of the CN line could be clearly
detectable. \textit{Line R21} is strongly blended with the CN line at
the same wavelength, hence its contribution may be detectable in the
core of the line. \textit{Line R15} is partially blended with a weaker
CN line at 2283.36 nm, but its left wing is clearly
visible. \textit{Line R14} is partially blended with a weaker CN line at
2289.27 nm, but its right wing allows the fluorine
determination. \textit{Line R16} is almost clean and quite weak, it is
the best fluorine indicator in this spectral range.  In the next
Section we will estimate the effect of the above CN blends on the
fluorine abundance together with the uncertainty coming from stellar
parameters.

\subsubsection{Uncertainties in open clusters \label{uncclu}}
The two main sources of uncertainty are the adopted stellar parameters
and the presence of CN blends.

In order to estimate the first one, we varied one stellar parameter at
a time leaving the others unchanged. We varied T$_{eff}$ by $\pm$150 K,
logg by $\pm$0.5 and $\xi$ by $\pm$1. For each new set of stellar
parameters we evaluated the new CNO abundances that well fit the spectrum.
Table \ref{tab_err} shows the fluorine variations and allows us to
estimate the fluorine abundance with different stellar parameters. In
particular using spectroscopic stellar parameters from \citet{mag10},
we obtain the following results: 
\textit{NGC6404-16}: log(F/H)$_{(spec)}$-log(F/H)$_{(this~study)}$ $\approx$~+0.2. \textit{NGC6404-27}: log(F/H)$_{(spec)}$-log(F/H)$_{(this~study)}$ $\approx$~+0.2. \textit{NGC6404-40}: log(F/H)$_{(spec)}$-log(F/H)$_{(this~study)}$ $\approx$~0. 

\begin{table}[ht]
\begin{center}
\caption{$\Delta$log(F/H) due to uncertainties in stellar parameters
and CN blends.\label{tab_err}}
\begin{tabular}{cccccccc}
\tableline\tableline
$\Delta$T$_{eff}$+150 & $\Delta$T$_{eff}$-150 & $\Delta$logg+0.5 & $\Delta$logg-0.5 & $\Delta\xi$+1 & $\Delta\xi$-1 & & \\
 (K)          &  (K)          &  dex     &    dex   &   $km~ s^{-1}$  &  $km~ s^{-1}$ \\
\tableline
+0.1 & -0.1 & -0.1 & +0.1  & 0.00 & 0.00 & & \\
\tableline\tableline
\multicolumn{8}{l}{Fluorine line-by-line variations due to CN blends} \\
\tableline
$\Delta$CN & R20 & R19 &R21 & R16 &R15 & R14& Average \\
\tableline
+0.1        & -0.05 & -0.02 & -0.25 & -0.02 & -0.07 & -0.02 & -0.07\\
\tableline
-0.1        & +0.05 & +0.02 & +0.25 & +0.02 & +0.07 & +0.02 & +0.07\\
\tableline
\end{tabular}
\end{center}
\end{table}

The influence of CN blends on the above F determination was studied in
the following way. We calculated two further synthetic spectra: one
with an enhanced (+0.1 dex) CN abundance; the second
with a lower (-0.1 dex) CN abundance. Then we derived the fluorine abundance and
results are shown in Table \ref{tab_err} (bottom). The first row
corresponds to results with an enhanced CN, while the second row refers to the analysis with the lower CN abundance. Each column shows
the fluorine abundance variation as derived from the corresponding HF
line, while the last column reports the average value of this
variation. Only the R21 line shows a large fluorine variation, but it
was expected because of the strong and central blend of this line with
a CN feature. Finally, we
estimated that the CN uncertainty influences the fluorine abundance by
less than $\pm$ 0.07 dex.

\section{Results}
Given the new experimental HF line parameters we recommend
a downward revision of the solar fluorine abundance from the old value
of $A(F) = 4.56$ determined by \citet{hal69} to the new value of
$A(F) = 4.40 \pm 0.25$, where the error is based on the
uncertainty of the atmospheric model that best matches the temperature
structure of the observed sunspot umbra.

For the OCs we present the results of our analysis in Table
\ref{tab_res}. It shows the fluorine log(F/H) (with log(X/H)=
log(N(X)/N(H)) + 12) as derived from each line (column 2-7). The star
mean abundance is shown in column 8. The cluster mean fluorine abundance is also given. The uncertainty associated with the single line abundance
corresponds to the $\sigma$ of the synthesis. \textbf{This latter corresponds to the standard deviation of the distribution of the residuals between the original spectrum and our best synthesis. We calculated the residuals over a wide portion of the spectrum (not only along the HF lines). Hence, the standard deviation of the distribution includes also the uncertainty in the continuum placement and the point-to point scatter. Despite this, for the uncertainty in the cluster mean abundance, we prefer to adopt the star-to-star abundance scatter as our best estimate of the error in the analysis. We then added to this the uncertainties due to blends with CN lines and due to the determination of atmospheric parameters.} 
We note that the warmer stars in the two clusters show abundances systematically higher (by about 0.1 dex) with respect to the cooler giants S488 (in M 67) and 40 in (NGC 6404). This behaviour is motivated by the fact that in warmer stars, the fluorine detection is more difficult, HF lines are less visible and less sensitive to abundance variations. 
Although in the cluster mean abundances we attributed the same weight to all the analyzed stars, we consider the estimates in the two cooler stars as more reliable.    
\begin{table}[ht]
\footnotesize
\begin{center}
\caption{Fluorine abundances in OC giants \label{tab_res}}
\begin{tabular}{cccccccc}
\tableline\tableline
        &   &   &    Log(F/H)   &    & & &  Log(F/H)$_{(avg)}$  \\
Line    &  R20 &  R19  &  R21  &  R15  &  R14    &  R16    &      \\
\tableline
Star        &      &       &       &   &         &     &\\
M~67-S364    & 4.46$\pm$0.10     &  4.46$\pm$0.10     &       &  4.46$\pm$0.08 &  4.46$\pm$0.10       &   4.56$\pm$0.08  &  4.50$\pm$0.06 \\
M~67-S488    & 4.36$\pm$0.05 &  4.36$\pm$0.05 &  4.56$\pm$0.05 &  4.36$\pm$0.05 &  4.36$\pm$0.05   &  4.36$\pm$0.05 & 4.39$\pm$0.08\\
M~67-S978    &      &      &  4.56$\pm$0.10     &  4.56$\pm$0.10 & 4.56$\pm$0.10        &  4.46$\pm$0.10  &  4.54$\pm$0.10\\
M~67-S1250   &      &       &       &  4.46$\pm$0.10 &  4.56$\pm$0.07   &   4.56$\pm$0.07 & 4.53$\pm$0.10\\
\tableline
\multicolumn{8}{c}{M~67 mean abundance  = 4.49$\pm$0.20}  \\
\tableline
Star       &      &       &       &   &         &     &   \\
NGC~6404-16      &      &       &       &  4.46$\pm$0.08 &  4.66$\pm$0.08       &   4.66$\pm$0.08 &  4.59$\pm$0.12  \\
NGC~6404-27      &      &  $<$4.66     &       &  $<$4.66 &   4.66$\pm$0.08  & 4.56$\pm$0.10    &  4.61$\pm$0.07  \\
NGC~6404-40      &  4.56$\pm$0.10    &  4.46$\pm$0.05     &  4.66$\pm$0.08     &  4.46$\pm$0.05 &  4.46$\pm$0.05  &  4.46$\pm$0.05 & 4.51$\pm$0.08  \\
\tableline
\multicolumn{8}{c}{NGC~6404 mean abundance  = 4.57$\pm$0.20}  \\
\tableline
\end{tabular}
\end{center}
\end{table}

\section{Discussion \label{disc}}
Our new solar determination of the fluorine abundance represents a strong improvement with respect to the previous analysis by \citet{hal69}. In fact, we adopted for the first time a set of experimental molecular parameters, modern spectrosynthesis tools, recent models of stellar atmosphere, we considered the magnetic field and the Zeeman splitting and a modern solar abundance compilation. All of these issues led us to a better description of the observed sunspot spectrum (e.g., good fit of all the CO, OH and HF features) and to a downward revision of the solar fluorine abundance, that now turns out to be A(F)=4.40$\pm$0.25. The new experimental molecular data are the main responsible for this revision, as we showed in Section 3.4. The error mainly reflects the uncertainty in the temperature determination of the sunspot spectrum. 


We note that our new solar fluorine is in a very good agreement with the meteoritic value, A(F)=4.42 from \citet{lod09}.
Moreover, we find that the very good abundance agreement between our solar determination and those in the M 67 giants, is a further strong test for the reliability of our solar result. As we said in the introduction, the M 67 cluster shows chemical abundances very similar to the Sun for all the studied elements. Our analyzed giants have not modified their initial fluorine abundance, therefore we expect, and actually find the same agreement between the M 67 and solar fluorine abundance, as for the other elements. Moreover, the analysis that we performed in its giant stars is completely different from that required in the Sun (e.g., no magnetic field, different spectrosynthesis code). Essentially, for the same metallicity, we made here two independent, different and careful analyses of fluorine, one in M 67 and one in the Sun. This results in a very good agreement and makes our estimate quite robust.

Considering fluorine in OCs, in a recent paper by \citet{nau13}, F abundance was determined in the Hyades, NGC 752, and M 67 clusters. Their different set of molecular parameters and solar fluorine abundance makes a comparison of their work with our results very difficult. A fluorine enhancement at younger ages is found by \citet{nau13}, but they used several upper limits for the  fluorine abundance. With our two determinations of fluorine in M 67 and NGC 6404 we can also contribute to trace the trend of the fluorine abundance with age. Indeed, the two clusters have very different age: 4.3 Gyr M 67; 0.5 Gyr NGC 6404 (see Table \ref{tab_clupar}). 

\textbf{We notice that the cloud from which NGC 6404 formed, received the yields of the nucleosynthesis of long-lived low-mass AGB stars (mass below $\sim$1.5 M$_{\odot}$). On the contrary, stars less massive than $\sim$1.5 M$_{\odot}$ did not have time to contribute to the chemical composition of the old M 67 primordial cloud. Therefore differences in the fluorine abundance in these OCs could be the signature of nucleosynthesis in the above low mass AGB stars. Recent works on stellar theoretical models \citep[see][]{mai12,tri14} suggest that low mass AGB stars should have a larger $^{13}$C-$^{14}$N pocket than previously assumed. These works explored the influence of a larger reservoir rich in neutron capture products and their parents. Since $^{14}$N is responsible for the fluorine production, an increase of its content inside low mass AGB stars could affect their fluorine production. The observation of the fluorine abundance in several open clusters with different ages, started in this work, will allow us: i) to trace the evolution of fluorine during the last 5-6 Gyr up to very recent periods; ii) through chemical evolution models, to estimate the contribution from low mass AGB stars; iii) to compare the latter with the prescriptions of the quoted new theoretical models.} 
 Looking at our cluster mean abundances, we find that the [F/H] ratio is slightly overabundant in the younger open cluster NGC 6404. The difference between the two OCs is however only $\sim$0.1 dex, well below the uncertainty of each measure, hence we are not yet able to provide convincing constraints to the fluorine evolution. Moreover, the two OCs are located at very different R$_{GC}$ and have a different metallicity. Indeed looking at the [F/Fe] ratios in the two analyzed OCs, we find solar [F/Fe] values in both, that in turn would indicate that the fluorine evolution is not strongly influenced by low-mass stars.

To summarize, new measurements of fluorine in many other open clusters are required. In particular, determinations in several OCs at different bins of Galactocentric radius and age are needed, in order to trace the fluorine evolution in different zones of the Galaxy. This would enhance the statistics of the analysis and would produce a more robust result even in this case, where individual abundance determinations are affected by uncertainties of around $\pm$0.2 dex.         

\section{Conclusions}

\begin{itemize}
\item We derived a new solar fluorine abundance in the spectral atlas
of \citet{wal01} of a medium strong sunspot umbra. We used experimental molecular data from the HITRAN database for the HF lines and modern spectrosynthesis tools, taking into account the magnetic field of the sunspot: our result is A(F)$_{\odot}$=4.40$\pm$0.25. 

\item We collected new spectra in the infrared region, with the CRIRES
  spectrograph, for 7 giant stars of two open clusters: M~67 and
  NGC~6404.

\item We derived fluorine abundances for the observed stars using: i)
  stellar parameters derived with photometric calibrations, and ii) the synthesis of their
  spectra. Uncertainties due to CN blends and stellar parameters were evaluated. The total error was estimated to be $\pm$0.20 dex. The abundance in M 67 is in a very good agreement with our new solar estimate, while fluorine in the younger OC NGC 6404 is $\sim$0.1 dex higher than the value in M 67. Looking at [F/Fe] ratios, we found solar values in the two analyzed OCs. 

\item Future studies of fluorine in several other open clusters with different ages and located at different R$_{GC}$, will allow us to trace its evolution in different zones of the Galaxy. It will also show the relevance of the AGB contribution to the synthesis of fluorine, improving our understanding of the its origin.
\end{itemize}

\acknowledgments
S.U. acknowledges support from the Austrian Science Fund (FWF) under project P22911-N16. \textbf{We would like to thank the referee for the useful suggestions and comments, that allowed us to improve this manuscript.}

\clearpage

\end{document}